\documentclass{aa} 

\usepackage{graphicx}
\usepackage{txfonts}
\usepackage{threeparttable}

\def\apj{{ApJ}}
\def\mnras{{ MNRAS}}

\def\araa{{ARA\&A}}

\def\be{\begin{equation}}
\def\ee{\end{equation}}
\def\bea{\begin{eqnarray}}
\def\eea{\end{eqnarray}}
\def\J{HESS\,J1858+020}
\def\G{SNR\,G35.6$-$0.4}
\newcommand{\Fermi}{{\sl Fermi}}
\def\HI{{\sc Hi}}
\def\HII{{\sc Hii}}

\begin{document}

   \title{HESS J1858+020: A GeV-TeV source possibly powered by CRs from SNR G35.6-0.4}

   \subtitle{}

   \author{Y. Cui\inst{1}, Y. Xin\inst{2}, S. Liu\inst{3}, P.H.T. Tam\inst{1}, G. P\"uhlhofer\inst{4}, and H. Zhu\inst{5}    }

\institute{School of Physics and Astronomy, Sun Yat-Sen University, Guangzhou, 510275, China \\
              \email{cuiyd@mail.sysu.edu.cn, tanbxuan@mail.sysu.edu.cn}
         \and
              School of Physical Science and Technology, Southwest Jiaotong University, Chengdu 610031, People's Republic of China
             \email{ylxin@swjtu.edu.cn}
       \and
              Key Laboratory of Dark Matter and Space Astronomy, Purple Mountain Observatory, Chinese Academy of Sciences, Nanjing 210033
                     \and
                     Institut f{\"u}r Astronomie und Astrophysik, Eberhard Karls Universit{\"a}t T{\"u}bingen, Sand 1, D 72076 T{\"u}bingen, Germany
                     \and
                     Key Laboratory of Optical Astronomy, National Astronomical Observatories, Chinese Academy of Sciences, Beijing 100101, China
             }

   \date{Received 28/07/2020; accepted 22/12/2020}

 
  \abstract
   {The supernova remnant (SNR) G35.6$-$0.4 shows a non-thermal radio shell, however, no $\gamma$-ray or X-ray counterparts have been found for it thus far. One TeV source, \J,~was found near the SNR and this source is spatially associated with some clouds at 3.6\,kpc. }
   {To attain a better understanding of the origin of \J, we further investigate the association between SNR cosmic rays (CRs) and the clouds through the \Fermi-LAT analysis and hadronic modeling.}
   { We performed the \Fermi-LAT analysis to explore the GeV emission in and around the SNR. We explored the SNR physics with previously observed multi-wavelength data.  We built a hadronic model using runaway CRs of the SNR to explain the GeV-TeV observation. }
   {We found a hard GeV source (SrcX2) that is spatially coincident with both \J~and a molecular cloud complex at 3.6\,kpc. In addition, a soft GeV source (SrcX1) was found at the northern edge of the SNR. The GeV spectrum of SrcX2 connects well with the TeV spectrum of \J. The entire $\gamma$-ray spectrum ranges from several GeV up to tens of TeV and it follows a power-law with an index of $\sim$2.15.
{We discuss several pieces of observational  evidence to support the middle-aged SNR argument.} Using runaway CRs from the SNR, our hadronic model explains the GeV-TeV emission at \J,~with a diffusion coefficient that is much lower than the Galactic value.}
   {}


\authorrunning{Cui et al.} \titlerunning{\Fermi-LAT analysis \& hadronic modeling of \J}

   \maketitle

%

\section{Introduction}

  \J~is one of the first unidentified TeV source reported by \cite{Ah2008}. In the HESS Galactic plane survey (HGPS) \citep{HESS2018}, \J~is also considered as one of eleven HGPS sources that do not yet have any associations with known physical objects. In the context of the HAWC telescope ($>100\,$GeV), which is mostly sensitive at $\sim$10\,TeV, there is no known 2HWC catalog source associated with HESS J1858+020 \citep{HAWC2017}.

The nearby radio source G35.6$-$0.4 located northwest of \J~was identified as a SNR by \cite{Green2009}. 
The radio boundary of this SNR is clearly shown in the 1.4GHz image with VGPS \citep{Stil2006,Zhu2013} and the 610 MHz image with GMRT \citep{Paredes2014}. 

A molecular cloud (MC) complex composed of two clumps has been found to be spatially coincident with \J. \cite{Paron2010} argued that the TeV emission of \J~is likely due to the interaction between the SNR CRs and this MC complex. This argument is further corroborated in \cite{Paron2011} by excluding the possibility of a young stellar object as the power source.
The velocity of this MC complex is $\sim55\mathrm{km\,s^{-1}}$, corresponding to a near-side distance of $\sim3.4\,$kpc or a far-side distance of $\sim10.4\,$kpc. The \HI~study by \cite{Zhu2013} suggested a distance to this SNR-cloud system of 3.6$\pm 0.4$\,kpc and this distance is also supported by the further \HI~study by \cite{Ranasinghe2018}. It was suggested that two nearby pulsars -- PSR J1857+0212 and PSR J1857+0210 could be be associated with the SNR \citep{Phillips1993,Morris2002}. However, these pulsar associations were disfavored by this newly confirmed distance of $3.6\pm0.4$ kpc because the dispersion measure distances to PSR J1857+0212 and PSR J1857+0210 were estimated to be 7.98\,kpc and 15.4\,kpc, respectively \citep{Han2006,Morris2002}.

No diffuse X-ray emission has been found in or around \G~with a 30\,ks Chandra observation \citep{Paredes2014}. However, seven X-ray point sources without IR counterparts have, in fact, been detected. Four of these point sources lie inside the SNR and their spectral properties resemble those of embedded protostars \citep{Paredes2014}. 

The previous \Fermi-LAT study on \G~by \cite{Torres2011} found no GeV detection. Hadronic models using the SNR as the power source were considered by \cite{Torres2011} and the authors suggested a low diffusion coefficient in order to explain the lack of GeV emission. In the 3FHL catalog, an extended source, 3FHL J1857.7+0246e, was shown to lie north of \J. However, the relation of 3FHL J1857.7+0246e, which is itself centered within the HESS J1857+026 region, with \J~is not straightforward. Recently, two point 4FGL sources at \G~were added in the newest 4FGL catalog \citep{Fermi2020}. Therefore, we proceeded to perform our own Fermi analysis of the SNR.

In this work, we further explore the hadronic origin of the GeV-TeV emission at \G. {In Section~\ref{Fermi_analysis}, we reanalyse the \Fermi-LAT data of \G~with improved data and tools. We find a hard GeV component at \J,~which has a smooth correlation with the HESS data, and its entire $\gamma$-ray spectrum extends from several GeV up to tens of TeV. Assuming the GeV-TeV emission of \J~is powered by escaped CRs from \G, then the scenario of a middle-aged SNR is favored because of its capacity to release low-energy CRs ($\lesssim$100\,GeV).  In Section~\ref{Multi-wavelength}, we look for evidence of a middle-aged SNR from previous multi-wavelength observations. In Section~\ref{SrcX2}, we build a hadronic model to explain the GeV-TeV spectrum of \J. }


\section{\Fermi-LAT data analysis}
\label{Fermi_analysis}
\subsection{Data preparation}
In the following analysis, we selected the latest \Fermi-LAT Pass 8 data with a ``Source'' event class (evclass=128 \& evtype=3), 
taken in the period between August 4, 2008 (Mission Elapsed Time 239557418) to May 1, 2019 (Mission Elapsed Time 578361605).
{For the spectral analysis, the energy range adopted is 1-500 GeV, while the analysis of the morphology was carried out in the 5-500 GeV range in consideration of the improved LAT resolution at higher energies.} The events with zenith angles greater than $90^\circ$ are excluded to reduce the contamination from the Earth Limb.
The region of interest (ROI) is a $14^\circ \times 14^\circ$ square region centered at the position of HESS J1858+020 \citep{Ah2008} and the standard LAT analysis software, {\em FermiTools,}\footnote {http://fermi.gsfc.nasa.gov/ssc/data/analysis/software/} was adopted.
The Galactic and isotropic diffuse background emissions are modeled according to {\tt gll\_iem\_v07.fits} and {\tt iso\_P8R3\_SOURCE\_V2\_v1.txt}\footnote {http://fermi.gsfc.nasa.gov/ssc/data/access/lat/BackgroundModels.html}.
All the sources listed in the 4FGL catalog \citep{Fermi2020} within a radius of $20^\circ$ from the ROI center, together with the two diffuse backgrounds, are included in the background model.

In the vicinity of HESS J1858+020, there are two 4FGL point sources called 4FGL J1858.3+0209 (R.A.=284.58$^\circ$, Dec.=2.15$^\circ$) and 4FGL J1857.6+0212 (R.A.=284.42$^\circ$, Dec.=2.20$^\circ$), {and yet these two 4FGL sources have no associations.}
Further to the north, also there is an extended source known as 4FGL J1857.7+0246e \citep{Fermi2020}, 
which is associated with HESS J1857+026 \citep{Ah2008}. Recent MAGIC observations revealed that the $>1$\,TeV emission at HESS J1857+026 can be spatially separated into two sources: MAGIC J1857.2+0263 and MAGIC J1857.6+0297 \citep{MAGIC2014}. The extension of 4FGL J1857.7+0246e is much larger than the TeV extension of HESS J1857+026 recorded by \citet{MAGIC2014}. 

Close to the SNR, {we can find the TeV source  \J. This nearly point-like source shows a slight extension of $5'$ \citep{Ah2008}, which is shown as a white circle in Fig.~\ref{fig:tsmap}.  The observations with MAGIC  \citep{MAGIC2014} also show a point-like source at \J~(named as MAGIC-south).} However, the TeV emission around the SNR is not investigated by the \cite{MAGIC2014} and no flux or spectral information of MAGIC-south is given due to its relatively low exposure at its angular distance from the MAGIC pointing positions. {In our \Fermi-LAT~analysis below, we also performed an analysis with the HESS/MAGIC image of \J~as an extended template. } 

\subsection{Spatial analysis of the GeV sources}
\label{Fermi_spatial}
\begin{figure}
\centering
\includegraphics[width=8.cm]{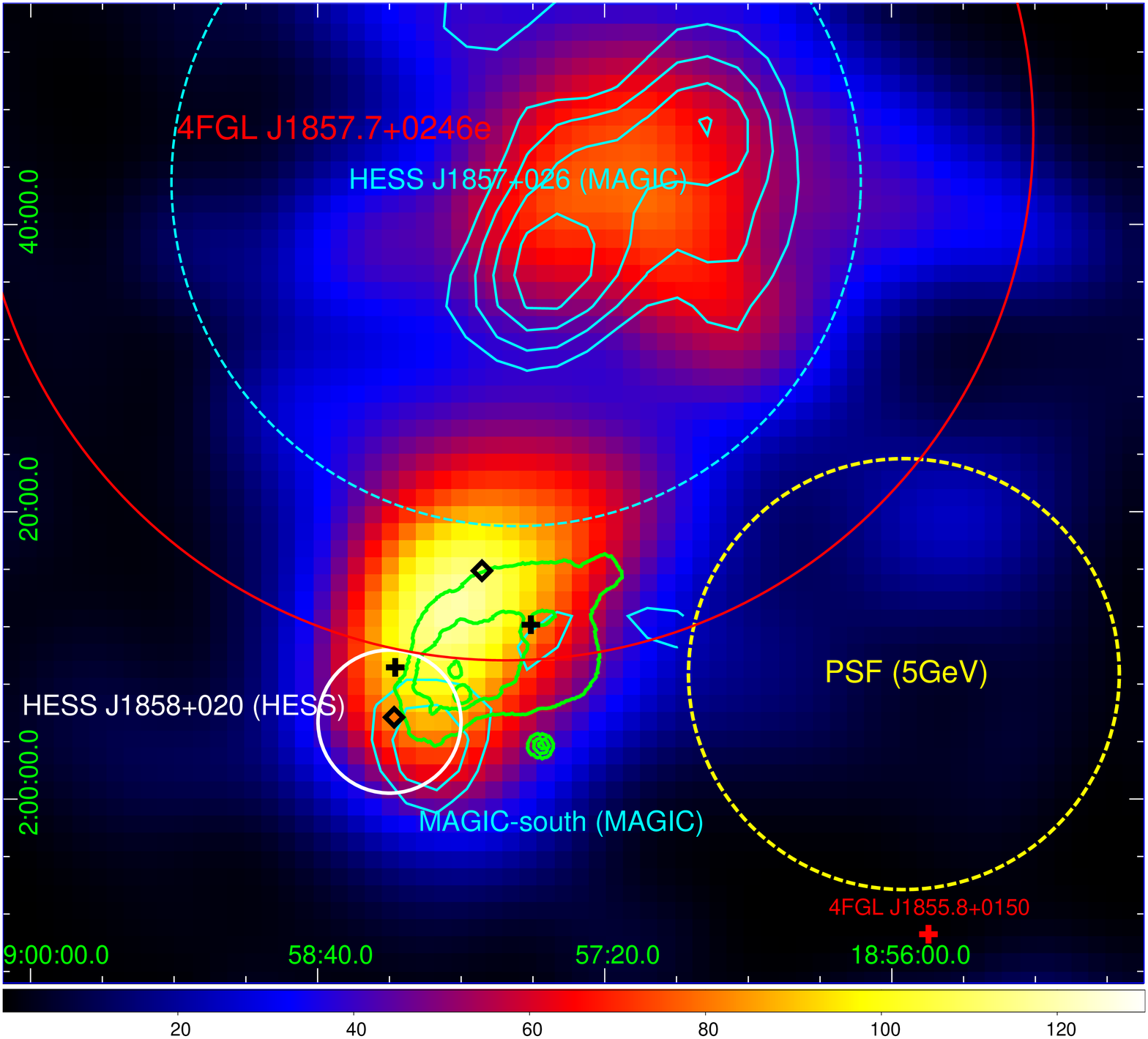} 

\includegraphics[width=8.cm]{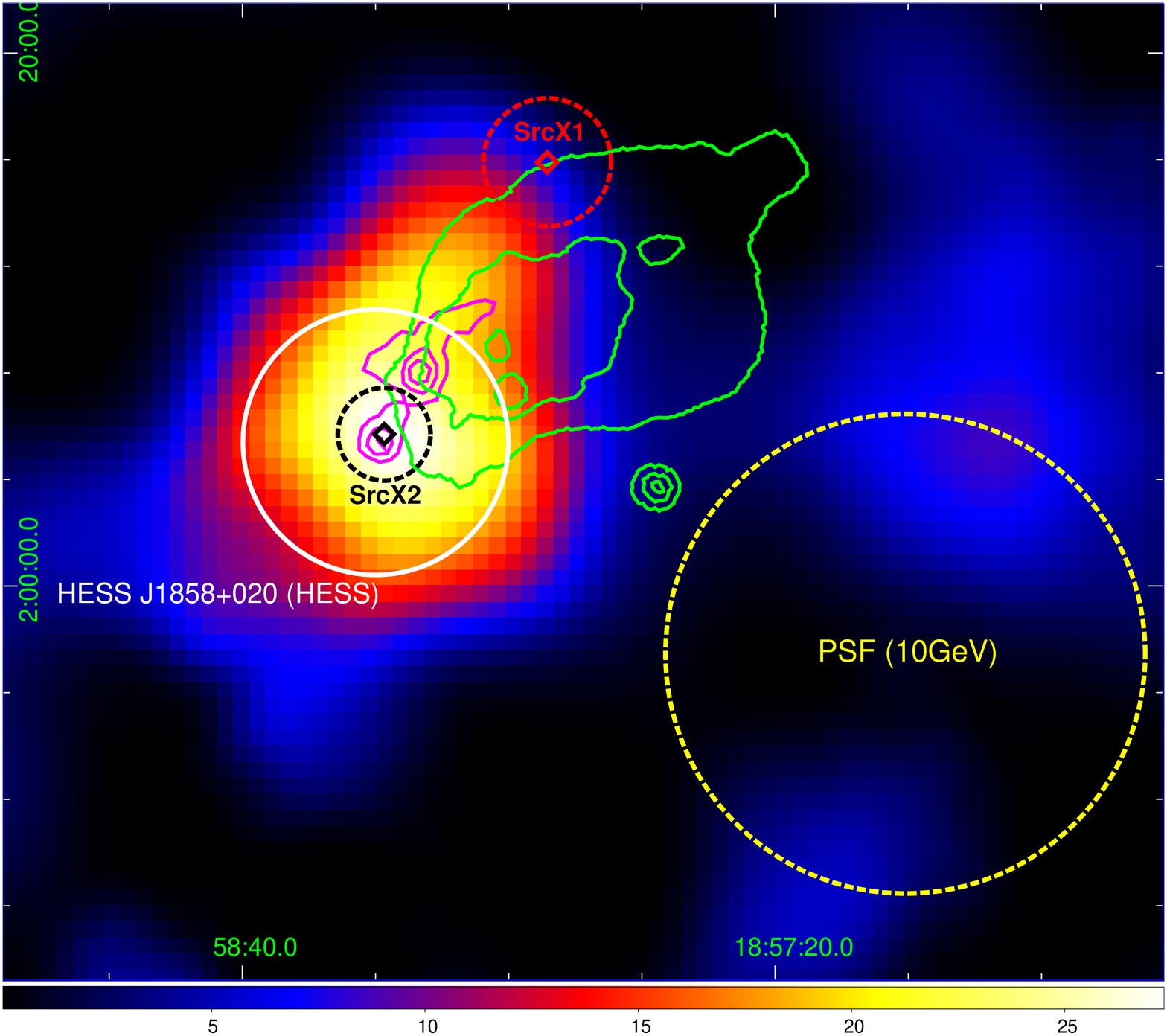}
\caption{{TS maps for photons above 5 GeV(top) and 10 GeV(bottom) are shown. Both of the maps} are smoothed with a $\sigma$=$0.03^\circ$ Gaussian kernel. {The PSF of a single 5 GeV/10 GeV photon is also shown in dashed yellow circles.} The red cross and circle show {a point source -- 4FGL J1855.8+0150  and an extended source -- 4FGL J1857.7+0246e, respectively.  The two black crosses represent two point sources -- 4FGL J1858.3+0209 and 4FGL J1857.6+0212.} The diamonds represent ``SrcX1" and ``SrcX2," which are the best fitted position of the two sources discovered in our analysis. {In the bottom panel, the 1$\sigma$ error radius of SrcX1 and SrcX2 are also shown in dashed circles.} The cyan dashed circle in the top panel shows the TeV extension of HESS J1857+026, which is adopted to be one of the spatial templates.  The cyan contours and {white circle} represent the MAGIC image and the HESS image, respectively.  The green contours show the VGPS data at 1.4 GHz with an arcminute resolution, and the magenta contours display the $^{13}$CO J = 1-0 emission for CloudX2. In both panels, the x-axis are the Right ascension (RA), and the y-axis are the Declination (Dec).}
\label{fig:tsmap}
\end{figure}

{For the purposes of carrying out a more detailed study of the emission around SNR G35.6-0.4/HESS J1858+020, we firstly derive test-statistic (TS) maps by excluding 4FGL J1858.3+0209, 4FGLJ1857.6+0212, and 4FGL J1857.7+0246e from the background model. 
Both the $>5\,$GeV the $>10\,$GeV TS maps are shown in Fig \ref{fig:tsmap}. As clearly seen in the SNR region, the 5\,GeV photons are mostly concentrated at the northeastern edge of the SNR, while the 10\,GeV photons are mostly around \J. 
This spatial difference between 5\,GeV and 10\,GeV TS maps seems to indicate a two source scenario.
Assuming these two sources are point sources, we hereafter fit the coordinates of the two sources with the command $\tt gtfindsrc$ and rename them as SrcX1 and SrcX2.}

{In the course of the following analysis, which focus on finding the spatial and spectra information of SrcX1 and SrcX2, the extended source 4FGL J1857.7+0246e was treated as a background source.
Both a uniform disk (the TeV extension of HESS J1857+026, the cyan circle in Fig.~\ref{fig:tsmap}) and the MAGIC image of HESS J1857+026 (the cyan contours north of the SNR in Fig.~\ref{fig:tsmap}) were used as the spatial template of 4FGL J1857.7+0246e and their analysis results in the scope of the SNR region essentially remain the same.}

Using only the $>5$\,GeV data, the best-fit position of SrcX1 was found to be R.A.$=284.476^\circ$, Dec.$=2.265^\circ$, with 1$\sigma$ error radius of 0.040$^\circ$. And for SrcX2, the best-fit position and its 1$\sigma$ error radius are R.A.$=284.578^\circ$, Dec.$=2.095^\circ$ and  0.029$^\circ$, respectively. The TS \citep{Mattox1996} values of SrcX1 and SrcX2 were fitted to be 49.1 and 39.3, corresponding to the significant level of 6.2$\sigma$ and 5.4$\sigma$, respectively. More comparisons between SrcX1 and SrcX2 in different energy bands are list in Table~\ref{table:Fermi}.


\begin{table*}
\centering
\caption {Best-fit position, spectral parameters, and TS values of SrcX1/SrcX2 for different energy bands with point source assumptions}
\begin{tabular}{cccccc}
\hline \hline
5-500 GeV   & R.A. \& Dec. & 1$\sigma$ error    & Spectral  & Photon Flux                        & TS         \\
            &           & radius             &  Index    & ($10^{-10}$ ph cm$^{-2}$ s$^{-1}$) & Value      \\
\hline
SrcX1       & $284.476^\circ$ \& $2.265^\circ$  & $0.040^\circ$     & $3.73\pm0.49$  & $2.62\pm0.49$    & 49.1     \\
SrcX2       & $284.578^\circ$ \& $2.095^\circ$  & $0.029^\circ$     & $2.31\pm0.27$  & $1.99\pm0.45$    & 39.3     \\
\hline \hline
5-10 GeV    & & & & &      \\
\hline
SrcX1       & $284.452^\circ$ \& $2.251^\circ$       & $0.038^\circ$     & $3.02\pm0.94$  & $2.54\pm0.47$    & 48.6     \\
SrcX2       & $284.578^\circ$ \& $2.095^\circ$(fixed)  &              & 2.31(fixed)      & $1.15\pm0.39$    & 13.3     \\
\hline \hline
10-500 GeV  & & & & &      \\
\hline
SrcX1       & $284.476^\circ$ \& $2.265^\circ$(fixed)  &           & 3.73(fixed)      & <0.37    & 0.2     \\
SrcX2       & $284.554^\circ$ \& $2.076^\circ$       & $0.037^\circ$   & $2.30\pm0.41$  & $0.85\pm0.22$        & 28.2     \\
\hline \hline
\end{tabular}
\label{table:Fermi}
\end{table*}

{Additionaly, extended spatial templates of SrcX1 \& SrcX2 were also explored to test their spatial extensions. These templates include uniform disks with a different radius, the MAGIC-south image of HESS J1858+020 and the HESS image of HESS J1858+020. Ultimately, these extended spatial templates basically show no improvement in the significance of SrcX1/SrcX2 and point sources are enough to describe their $\gamma$-ray emissions.}

In Fig \ref{fig:tsmap}, the radio contours of \G~at 1.4GHz with VGPS \citep{Stil2006, Zhu2013} and the contours of nearby TeV sources \citep{Ah2008,MAGIC2014} are plotted. 
The eastern cloud traced by the $^{13}$CO J = 1-0 emission with a velocity range between 51 and 59 km s$^{-1}$ \citep{Paron2010} is shown in purple contours and named "CloudX2" for the purposes of this work.
As can be seen, the position of SrcX2 is in good correspondence with the TeV image of HESS J1858+020 and CloudX2, while SrcX1 appears to be corresponding to the radio shell of \G. 

\subsection{Spectral analysis of the GeV sources}
\label{Fermi_spectral}
With the best positions of SrcX1 and SrcX2 obtained from the 5-500 GeV data (see Table~\ref{table:Fermi}), we used {\tt gtlike} to fit the power-law spectra of them in the energy range of 1-500 GeV. 
The spectral index and total photon flux of SrcX1 are $3.09\pm0.09$ and $(6.55\pm0.52)\times10^{-9}\, \mathrm{photon\,cm^{-2}\,s^{-1}}$.
While the spectral index and total photon flux of SrcX2 are fitted as $2.27\pm0.14$ and $(1.59\pm0.40)\times10^{-9}\,\mathrm{photon\,cm^{-2}\,s^{-1}}$. 

\begin{figure}
\centering
\includegraphics[width=\columnwidth]{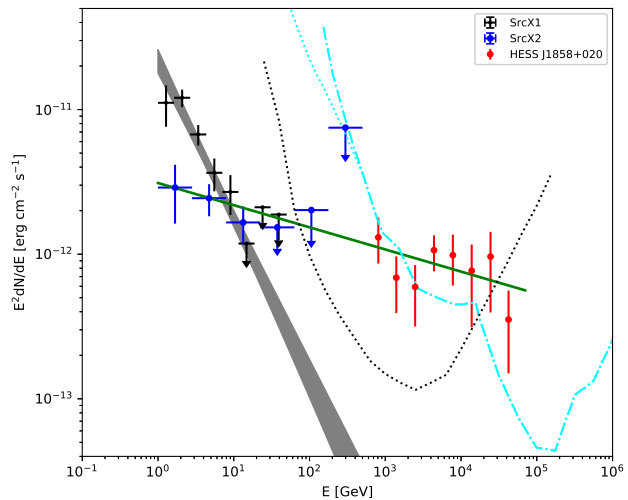}
\caption{\Fermi-LAT SED of SrcX1 (black dots) and SrcX2 (blue dots), with
arrows indicating the 95\% upper limits. Red dots represent the HESS observation of HESS J1858+020 \citep{Ah2008}. Gray butterfly indicates the best-fit power-law of SrcX1 in the energy range of 1-500 GeV. Solid green line is the joint fit for the \Fermi-LAT data of SrcX2 and HESS data of HESS J1858+020.
The cyan dotted and dot-dashed lines show the differential sensitivities of LHAASO (1 year) with different sizes of photomultiplier tube \citep[PMT;][]{Bai2019}. Black dotted line represents the differential sensitivity of CTA-North \citep[50 hrs;][]{CTA2019}.}
\label{fig:sed}
\end{figure}

To derive the spectral energy distribution (SED) of SrcX2 at different energies, 
we binned the data with six logarithmically even energy bins between 1 GeV and 500 GeV 
and we performed the same likelihood fitting analysis to the data. 
Considering the much softer spectrum of SrcX1, we derived the energy band of 1-50 GeV into eight bins to obtain its SED. 
For the energy bin with the TS value of SrcX1/SrcX2 smaller than 5.0, an upper limit with 95\% confidence level was calculated. 
The results of the spectral analysis are shown in Fig \ref{fig:sed}.

In summary, we performed the \Fermi-LAT analysis around \G~and found two distinct GeV sources:\ one displaying a hard GeV spectrum (SrcX2) is in spatial coincidence with \J~and CloudX2; the other one displaying a soft GeV spectrum (SrcX1) is located at the northern edge of the SNR. The GeV spectrum of SrcX2 connects well with the TeV spectrum of \J\ and together, they show a hard GeV-TeV spectrum with a power-law index of $\sim$2.15.

\section{Multi-wavelength observations around \G}
\label{Multi-wavelength}
\subsection{SNR size and distance}
\begin{figure}
\centering
\includegraphics[width=\columnwidth]{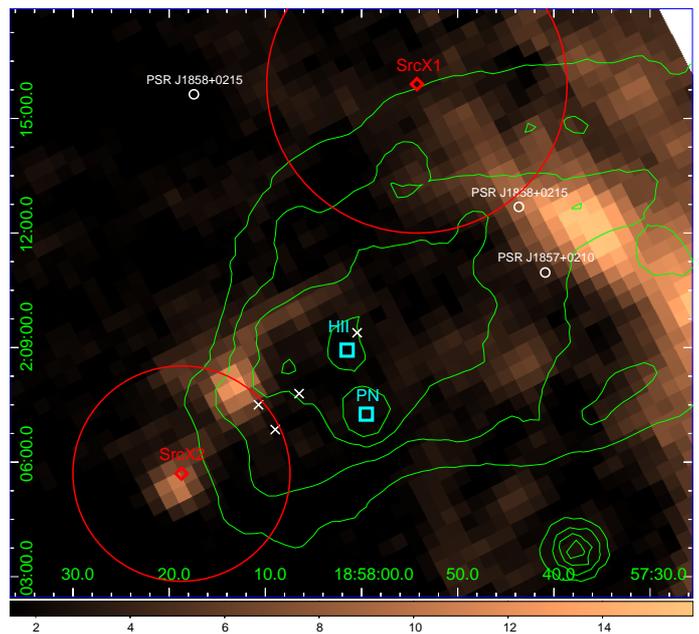}
\caption{ $^{13}$CO map around \G~is shown, where the CO data is obtained from GRS. The x-axis and y-axis represent the RA and Dec, respectively. {Noticeably, only the contours of CloudX2 is shown in Fig.~\ref{fig:tsmap}, meanwhile the western clouds that lack $\gamma$-ray counterparts are neglected in this work.} The 1.4GHz image from VGPS is marked in green contours. The best fitted position of SrcX1 and SrcX2 are marked in red diamonds, red circles represent their position uncertainties (2$\sigma$). The X-ray point sources and the nearby pulsars are marked in white crosses and circles, respectively. The \HII~region G35.6$-$0.5 and the planetary nebulae PN\,G35.5$-$0.4 are marked in cyan boxes.}
\label{fig:CO}
\end{figure}

In adopting the 1.4GHz image with VGPS (see Fig.~\ref{fig:CO}), we obtained a close-up of the complex with an angular size of $\sim13' -17'$ and a center at RA$=284.457^\circ$, Dec$= 2.176^\circ$ \citep{Zhu2013}. 

The most up-to-date \HI~\& CO studies by \cite{Zhu2013} and \cite{Ranasinghe2018} suggest that the distance to the SNR-cloud system (\G~\&~Eastern cloud) is 3.6$\pm 0.4$\,kpc, and 3.8$\pm 0.3$\,kpc, respectively.  We adopted a distance of 3.6\,kpc in our model and this value leads to an averaged SNR radius of $\sim8$\,pc.

\subsection{Considering a possible middle-aged SNR}
{As described in Section~\ref{Fermi_spectral}, the $\gamma$-ray spectrum at CloudX2 extending down to several GeV seems to indicate that this is the case of a middle-aged SNR.}
 So far, there is no direct observational evidence for the shock velocity.  {In following: we discuss the evidence based on multi-wavelength studies} in favor of a middle-aged SNR.
 
{\it$\gamma$-ray emission extending down to several GeV: } The flat $\gamma$-ray spectrum at CloudX2 ranges from several GeV to tens of TeV, which is similar to those of the middle-aged SNRs associated with MCs, such as SNR W28, W44, and IC443. Its flat spectral shape, in particular, resembles those of the clouds (240A,B,C) next to but not in direct proximity to SNR\,W28 \citep{Abdo2010,Hanabata2014,Cui2018}. If a hadronic origin is assumed, then the SNR has already released CRs with energies down to tens of GeV into CloudX2. A slow shock ($v_\mathrm{SNR}\ll1000\mathrm{km\,s^{-1}}$) at present can achieve an escape energy of $\sim10$\,GeV with the help of the damping of magnetic waves by neutrals \citep{Drury1996}. It could be argued that a young SNR can also release the GeV CRs to MCs through a shock-cloud collision scenario \citep{Cui2019,Tang2015}. Nonetheless, {there is no evidence from the linewidth of the molecular clouds that are interacting with the SNR; for example,  no line asymmetry was found at CloudX2 \citep{Paron2011}.} \\

{\it An intrinsic weak X-ray emission}: The non-detection of diffuse X-ray emission \citep{Paredes2014} can be either due to the intrinsic nature of the SNR or the heavy absorption.  The total H column density {along the line of sight (LOS) of G35.6$-$0.4 is $N_\mathrm{H,total}\approx1.44\times10^{22}\,\mathrm{H\,cm^{-2}}$ \citep{Willingale2013}. }The extinction curve along the LOS of the SNR~becomes flat behind {$\sim3.4$\,kpc \citep{Green2019}. This indicates that most of the observed gas lies in front of the SNR (3.6\,kpc) and the column density up to 3.6\,kpc is about $96\% N_\mathrm{H,total}$.} Obviously, this $1.44\times10^{22}\,\mathrm{H\,cm^{-2}}$ is not thick enough to absorb most keV photons; see, for example, the SNRs with higher $N_\mathrm{H}$ displaying X-ray emissions \citep{Zhu2017}. Hence, the non-detection of diffuse X-ray emission is likely due to an intrinsic weak source. 
\\

{\it A hard radio index:} The non-thermal radio spectrum of \G~displays an index of $\alpha = 0.47\pm0.07$ \citep{Green2009}. This value is much harder than the ones ($\alpha\approx 0.6-0.8$) measured in young SNRs \citep{Dubner2015}.

Ultimately, we find that \G~is likely to prove a middle-aged SNR but none of this evidence is conclusive at present. {Noticeably, with an ambient density of $1\,\mathrm{H\,cm^{-3}}$, a SNR distance of $3.6$\,kpc will lead to a SNR age of merely 2\,kyr. In the modeling section below, we adopt this middle-aged SNR scenario in order to improve the release of $<100$\,GeV CRs -- hence, a relatively higher ambient density is required. }

\subsection{The circumstellar medium and other observations} 

 Following the $^{13}$CO study by \cite{Paron2010} and the distance study by \cite{Zhu2013}, CloudX2 is shown to have a mass of $\sim5.0\times10^3M_\odot$; see also Fig.~\ref{fig:CO}. The projected distance between CloudX2 and the SNR center is $\sim8\,$pc. The even larger clouds located at the western side of the SNR lack GeV-TeV counterpart, hence, they are not considered in \cite{Paron2010} and they are not included in our Fig.~\ref{fig:tsmap} either. In our model, shown below, they are assumed to be located far from the SNR.

Noticeably, a \HI~shell sometimes could be associated with a pre-SN wind bubble or a SNR; see, for example, the slow-expanding \HI~shell around the Wolf-Rayet star HD 156385 \citep{Cappa1988} and the fast-expanding \HI~shells around {SNR CTB 80 \citep{Park2013} \& SNR Cygnus Loop \citep{Leahy2003}.  Finding such a shell could help us to confine the progenitor type and the ambient density, which eventually leads to a more accurate SNR age. Unfortunately, \HI~shells are very difficult to find and only several are known among all $\sim300$ Galactic SNRs \citep{Leahy2012}.} We revisited the \HI~data used in \cite{Zhu2013} and we do not find any evidence for either a slow or a fast \HI~shell around \G.

As seen in Fig.~\ref{fig:CO}, the GeV source SrcX1 lacks multi-wavelength counterparts and no X-ray source, TeV source, pulsar, or \HII~region is located within the $2\sigma$ uncertainty circle. Hence, SrcX1 will be considered as a background source in this work.

\section{A hadronic explanation of GeV-TeV emission of \J}
\label{SrcX2}
\subsection{Models}
As described in the previous section, one of the most challenging characteristics of the $\gamma$-ray spectrum of SrcX2 is the broad energy range. A successful acceleration model for such a spectrum should be able to release the $\gtrsim100$\,TeV CRs during its early SNR stage, as well as the $10-100$\,GeV CRs during its late SNR stage.

\begin{figure}
\centering
\includegraphics[width=\columnwidth]{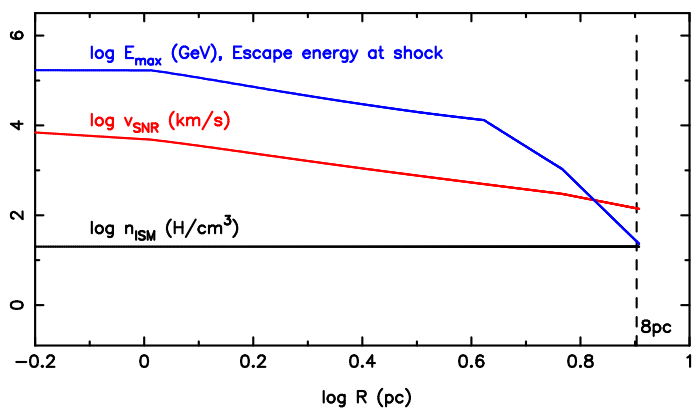} 
\caption{SNR evolution profiles. The radial profiles of the circumstellar density, shock velocity, and the escape energy are shown in black, red, and blue lines, respectively.}
\label{fig:SNR}
\end{figure}

A simple SNR evolution history with a homogeneous circumstellar medium is adopted in our model.
To calculate the SNR evolution history, {the analytical solution of $v_\mathrm{SNR} \propto t^{-3/7}$ \citep{Chevalier1982,Nadezhin1985} was adopted for the ejecta-dominated stage, while the thin-shell approximation \citep{Ptuskin2005} was adopted for the Sedov-Taylor stage; the result is essentially consistent with the analytical solution of $v_\mathrm{SNR} \propto t^{-3/5}$ \citep{Ostriker1988,Bisnovatyi1995} and the analytical solution of $v_\mathrm{SNR} \propto t^{-7/10}$ by \cite{Cioffi1988}} is adopted for the pressure-driven snowplow (PDS) stage. The SNR evolution profiles are shown in Fig.~\ref{fig:SNR}.

In explaining the hard TeV tail of SrcX2, we adopt the acceleration theory of nonresonant streaming instability developed by \cite{Bell2004} and \cite{Zirakashvili2008}. This theory can boost the escape energy up to hundreds of TeV in young SNRs.  Given a SNR's evolution history and an acceleration efficiency, an analytical approximation of this theory \citep{Zirakashvili2008} can provide us with the runaway CR flux $J$ and the escape energy $E_\mathrm{max}$. In a strong shock, only CRs with energies above $E_\mathrm{max}$ can escape from the shock upstream and become runaway CRs. A magnetic field of $B_0=5\mathrm{\mu G}$ and an initial magnetic fluctuation of $B_\mathrm{b}=7\%B_0$ in the ICM are assumed in the calculation, following \cite{Zirakashvili2008}.

In explaining the GeV spectrum of SrcX2, the damping of the magnetic waves by the neutrals is adopted in the late SNR stage. This damping effect can significantly lower $E_\mathrm{max}$ and it is considered important in mid-aged and old SNRs; both \cite{Shull1979} and \cite{Sutherland2017} noted that a shock slower than $\sim100\,\mathrm{km\,s^{-1}}$ will lead to a significant drop of the UV ionization at shock precursor. The relationship of $E_\mathrm{max}= v_\mathrm{3}^3n_\mathrm{H}^{0.5}n_\mathrm{n}^{-1}\,$TeV \citep{Drury1996} is adopted to estimate the escape energy in partially ionized medium, where $v_\mathrm{3}$ is shock velocity in unit of $10^3 \mathrm{km\,s^{-1}}$, $n_\mathrm{H}$ is the circumstellar density, and $n_\mathrm{n}$ is the neutral density. A homogeneous diffusion coefficient was adopted in the entire space, which follows a power-law rule of $D=D_{10}E^{\delta}$. By integrating the SNR surface as well as the entire SNR evolution history,  it is possible to obtain the present CR density at CloudX2; see also the equations in Section 2.4.1 of \cite{Cui2016}.

\subsection{Parameter justifications and results}
 The escape energy $E_\mathrm{max}$ is highly sensitive to the shock velocity, $v_\mathrm{SNR}$. To generate $\sim10\,$GeV CRs, one requires a shock velocity of $v_\mathrm{SNR}\ll 1000\,\mathrm{km\,s^{-1}}$ during the late stage of the SNR evolution. Hence, we adopt a relative high circumstellar density of $n_\mathrm{H}=20\,\mathrm{H\,cm^{-3}}$. {Such a density at a Galactocentric distance of 5\,kpc indicate that the circumstellar medium is cold neutral medium \citep{Wolfire2003,Cox2005}.}
 When the SNR reaches 8\,pc, our model gives a SNR age of $t_\mathrm{SNR}=$18\,kyr, a shock velocity of $v_\mathrm{SNR}=142\,\mathrm{km\,s^{-1}}$, and an escape energy of $E_\mathrm{max}=26\,$GeV, see Fig.~\ref{fig:SNR}. 
 
{More details of the best fitted parameters are shown in Table~\ref{table:SNR}, where we also show the dependencies of our fitting results on those parameters. A higher $E_\mathrm{ej}$ gives a overall higher $v_\mathrm{SNR}$. Both a higher $v_\mathrm{SNR}$ and a higher $\eta$ can eventually leads to a higher total CR production and a higher escape energy $E_\mathrm{max}$.  The value of $E_\mathrm{ej}$ is suggested to be around $1\,\rm{\mathcal{E}_{51}}$. The value of $\eta$ is limited by that the energy of total accelerated CRs should not be too far from $10\%E_\mathrm{ej}$.  A lower diffusion coefficient, $D,$ and longer SNR-cloud distance, $L,$ will suppress the CRs (mostly GeV CRs) from reaching CloudX2. However, once the CRs can easily reach the cloud (mean diffusion distance after certain time is beyond the SNR-cloud distance $L$), a lower diffusion coefficient helps confining the CRs in the SNR-cloud region from spreading too thin.
}

The neutral density $n_\mathrm{n}$ in the shock precursor lacks observational constraints. Following the recent simulation work (Fig.~5 in \cite{Sutherland2017}), we adopt an estimation of $ X=4.5\%\cdot(v_\mathrm{SNR}/100\,\mathrm{km\,s^{-1}})^{-2}$ in a range of $v_\mathrm{SNR}= 140 - 500\,\mathrm{km\,s^{-1}}$, where $X=n_\mathrm{n}/n_\mathrm{H}$ and $1-X$ is the ionization ratio. In a more detailed model, for a future study, $X$ is dependent on many other factors, such as metallicity or magnetic field; see also \cite{Sutherland2017}. The damping of magnetic waves takes effect when the shock velocity is below $\sim500\,\mathrm{km\,s^{-1}}$, and this $500\,\mathrm{km\,s^{-1}}$ is chosen to get a smooth transition from the theory of \cite{Zirakashvili2008} to that of \cite{Drury1996}.

\begin{table}
\caption{Parameters of the SNR model} 
\begin{threeparttable}           
\label{table:SNR}     
\begin{tabular}{c c c c c c c }     
\hline\hline       
  & $E_\mathrm{ej}$  $^{a}$    & $\eta$  $^{b}$ & $D_\mathrm{10}$ $^{c}$  & $\delta$ $^{c}$  & $L$  $^{d}$
 \\
\hline
parameter value & $1.2\,\rm{\mathcal{E}_{51}}$   & 0.08 & 0.24 & 0.25 &    10\,pc  \\
\hline
$\gamma$-ray flux $^{e}$ & $+$  & $+$ &  &  &  $-$  \\
\hline
spectral index $^{f}$ & $+$   & $+$ & $-$ & $-$ &   $+$  \\
\hline                  
\end{tabular}
 \begin{tablenotes}
     \item[a] Explosion energy of the SN. ${\mathcal{E}_{51}}=10^{51}\,$erg.
     \item[b] The acceleration efficiency $\eta$ represents the ratio between the energy flux of runaway CRs and the kinetic energy flux of incoming gas onto the shock, and it remains constant through the entire SNR evolution. The total energy of all the released CRs is $23\%\,E_\mathrm{ej}$.
     \item[c] The diffusion coefficient follows a rule of $D=D_{10}E^{\delta}$, where $D_{10}$ is in unit of $10^{28}\,\mathrm{cm^2\,s^{-1}}$. $D_{10}=1$ and $\delta=0.3-0.5$ represent the Galactic diffusion coefficient.
     \item[d] The three dimensional distance between the SNR and CloudX2 (The projected distance is $\sim8\,$pc).
     \item[e] {``+" / ``-" means that with a increasing value of the parameter, the $\gamma$-ray flux (total energy of GeV-TeV band) of SrcX2 increases/decreases. For parameter $D$, both ``+" and ``-" could happen, that is why we leave them blank, see also the explanations in text.}
     \item[f] {``+" / ``-" means that with a increasing value of the parameter, the $\gamma$-ray spectral index becomes harder/softer (higher/lower TeV to GeV ratio).}
   \end{tablenotes}
  \end{threeparttable}
\end{table}

\begin{figure}
\centering
\includegraphics[width=\columnwidth]{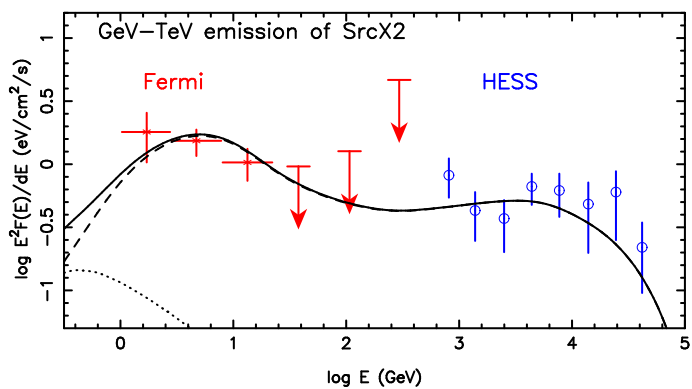} 
\caption{Hadronic model results using SNR CRs in explaining the GeV-TeV emission of SrcX2. The \Fermi-LAT data and HESS data of SrcX2 are marked in red and blue, respectively. Our model results are shown in solid lines, the contribution of runaway CRs and CR sea are marked in dashed lines and dotted lines, respectively.}
\label{fig:fitting_SrcX2}
\end{figure}

Using the nonresonate acceleration model in the early stage and the damping model in the late stage, over  the entire SNR evolution, the runaway CRs cover a large energy range, from $\sim$10\,GeV up to more than 100\,TeV, see Fig.~\ref{fig:SNR}. To explain the flat GeV-TeV spectrum of SrcX2, the early released TeV CRs should not diffuse too far at an age of 18\,kyr, meanwhile the late-released GeV CRs should be able to reach CloudX2. Hence, a relative hard index of diffusion coefficient (0.25) is adopted; see also Table~\ref{table:SNR}.
 Ultimately, these SNR CRs explain the GeV-TeV emission of SrcX2 with a diffusion coefficient that is much lower than the Galactic value; see the spectrum fitting in Fig.~\ref{fig:fitting_SrcX2}. The SNR has a Galactocentric distance of 5\,kpc, while the corresponding CR sea contribution \citep{Yang2016,Acero2016} is very little and can be ignored.

\section{Discussions and observational expectations}
\label{discussion}
 One of the arguments in support of \G~as a middle-aged SNR is the non-detection of diffuse X-ray emission with Chandra.
 More sensitive observations with XMM-Newton may solve this question by detecting either the thermal or the non-thermal emission.
For instance, the thermal emission found in SNR W28 with XMM-Newton \citep{Zhou2014} shows a temperature of $\sim0.5\,$keV with a column density of $\sim4\times10^{21}\,\mathrm{H\,cm^{-2}}$. The column density at \G~is $\sim1.44\times10^{22}\,\mathrm{H\,cm^{-2}}$. 
Future X-ray observations may also discover the alternative power sources if they are leptonic dominated -- for example, a PWN origin for HESS J1640-465 \citep{Xin2018} -- as well as the potential shock-cloud collisions -- for example, the thermal X-ray emission at the northeastern shell of SNR W28 \citep{Zhou2014}. Noticeably, millimeter observations of ionization lines may also shed some light on the shock-cloud collisions.

One of the most interesting feature of SrcX2 is the hard TeV tail and we expect future observations with LHAASO/CTA may further characterize the hard tail (see Fig.~\ref{fig:sed}). In addition, they may also reveal more detailed TeV features at or around the SNR.  

\section{Summary}

We carried out an analysis the \Fermi-LAT data at \G~and discovered two GeV sources using the $>5$\,GeV data. A soft GeV source -- SrcX1 is located at the northern edge of the SNR, a hard GeV source -- SrcX2 is in spatial coincident with \J~and the molecular cloud complex at east -- CloudX2. The spectral index of SrcX1 and SrcX2 are 3.09$\pm$0.09 and 2.27$\pm$0.14, respectively. The GeV spectrum of SrcX2 connects well with the TeV spectrum of \J. The entire GeV-TeV spectrum of SrcX2 is flat  and it covers a wide energy range, from several GeV up to tens of TeV.

We find that \G~is possibly a middle-aged SNR, and this argument is supported by three pieces of indirect observational evidence. Firstly, we find that the lack of diffuse X-ray emission, especially for the keV band, is likely due to an intrinsic weak source rather than the heavy absorption. Secondly, if the SrcX2 is indeed powered by the SNR CRs, then the GeV emission found at CloudX2 indicates that CRs with energies down to $\sim10\,$GeV have been released from the SNR. Thirdly, the radio index of \G~is much harder than that of a young SNR. However, this evidence is not conclusive and we look forward to future observations of \G.

 We built a hadronic model to explain the GeV-TeV emission of SrcX2 with SNR CRs. By adopting the acceleration theory of nonresonant streaming instability, our model can generate CRs with energies of $\gtrsim100$\,TeV during the early SNR stage. The damping of magnetic waves by the neutrals was adopted for the late SNR stage and it leads to the release of CRs with energies down to $\sim10\,$GeV.  Our model requires a diffusion coefficient that is much lower than the Galactic value  and,  in particular, a hard index of diffusion coefficients is needed to suppress the diffusion of early-released TeV CRs.

\begin{acknowledgements}
      We like to thank Guangxing Li, Wenwu Tian for discussions on radio studies of \G. Yudong Cui \& P.H. Thomas Tam are supported by the Fundamental Research Funds for the Central Universities grant (20lgpy170) and the National Science Foundation of China (NSFC) grants (11633007, 11661161010, and U1731136). Yuliang Xin is supported by the Natural Science Foundation for Young Scholars of Jiangsu Province, China (No. BK20191109). Siming Liu is supported by NSFC (No. U1738122, U1931204) and the International Partnership Program of Chinese Academy of Sciences (No.114332KYSB20170008). Hui Zhu is supported by National Key R\&D Program of China (2018YFA0404203) and NSFC (11603039).
\end{acknowledgements}

%
%

\end{document}